\documentclass{ws-procs961x669}            

\begin{document}
\title{Holographic Approaches to DIS on a Nucleus}

\author{Kiminad A. Mamo$^*$}

\address{Department of Physics and Astronomy, Stony Brook University,\\
Stony Brook, New York 11794-3800, USA\\
$^*$E-mail: kiminad.mamo@stonybrook.edu}

\begin{abstract}
We consider deep inelastic scattering (DIS) on a dense nucleus described as an extremal RN-AdS black hole with holographic quantum fermions in the bulk. We find that the R-ratio (the ratio of the structure function of the black hole to proton) exhibit  shadowing for $x < 0.1$, anti-shadowing for $0.1 < x < 0.3$, EMC-like effect for $0.3 < x < 0.8$ and Fermi motion for $x > 0.8$ in a qualitative agreement with the experimental observation of the ratio for DIS on nucleus for all range of $x$. We also take the dilute limit of the black hole and show that its R-ratio exhibits EMC-like effect for $0.2 < x < 0.8$ and the Fermi motion for $x > 0.8$, and no shadowing is observed in the dilute limit for both bottom-up (using Thomas-Fermi approximation for the nucleon distribution inside the dilute nucleus), and top-down (considering the dilute nucleus to be a Fermi gas in AdS) approaches.
\end{abstract}


\bodymatter

\section{Introduction}
We use holography or the AdS/CFT correspondence to model deep inelastic scattering (DIS) on both dilute and dense nuclei \cite{Mamo:2018eoy,Mamo:2018ync,Mamo:2019jia}. See Ref. 4 for application of holography in proton-nucleus collisions. We model a dilute nucleus in a bottom-up and top-down approaches. In the bottom-up approach, we consider a nucleus to be made of nucleons distributed according to the Thomas-Fermi approximation, and DIS on the dilute nucleus is dominated by the incoherent scattering on individual nucleons. Therefore, in this bottom-up approach, we use the structure function of each nucleons calculated in holography as input to determine the structure function of the nucleus. We find that the R-ratio exhibits an EMC effect in the large-x regime.

In the top-down approach, we consider the nucleus to be a Fermi gas in AdS spacetime. And, by carrying out the DIS directly on the Fermi gas, we determine the structure function of the nucleus from the current-current correlation function computed by using a one-loop Witten diagram in pure AdS where the fermions are running in the loop forming a Fermi gas. We find that the R-ratio exhibits EMC effect, similar to the bottom-up approach.

We model the dense nucleus by an extremal RN-AdS black hole, and identify the structure function of the extremal RN-AdS black hole to the structure function of a nucleus. At leading order in $1/N_c$ expansion, the structure function of the nucleus, is extracted from the current-current correlation function of the classical extremal RN-AdS black hole background without matter contribution. We find that the structure function of the classical extremal RN-AdS black hole is mostly at small-x, and exhibits shadowing.

At sub-leading order in $1/N_c$ expansion, the highly damped bulk Dirac fermions (protons) due to the background RN-AdS black hole, also contribute to the extremal RN-AdS black hole's or the dense nucleus's structure function. We find that the quantum correction contributes mostly at large-x regime.

\section{DIS on a Dilute Nucleus}\label{dilute}

In holography, Compton scattering on a single nucleon at the boundary maps onto
the scattering of a U(1) current onto a bulk Dirac fermion which, at large-x, is
dominated by s-channel exchange of bulk Dirac fermion resonances, while at small-x
the same scattering is dominated by the t-channel exchange of spin-j glueball
resonances, with the interpolating result for the structure function of the proton
(modeled as a bulk Dirac fermion) as\cite{Polchinski:2002jw}
\begin{equation}
\label{F2p}
F_2^p(x, q^2)=\tilde{\mathbb C}\left(\frac{m_N^2}{-q^2}\right)^{\tau-1}
\left(x^{\tau+1}(1-x)^{\tau-2}+{\mathbb C}\left(\frac{m_N^2}{-q^2}\right)^{\frac 12}\frac 1{x^{\Delta_{\mathbb P}}}\right)\,,
\end{equation}
with $mR=3/2$ or $\tau=\Delta-1/2=3$.
\subsection{Bottom-up Approach to Dilute Nucleus}
\begin{figure}[t]
\begin{center}
\begin{tabular}{cc}
\includegraphics[width=6cm]{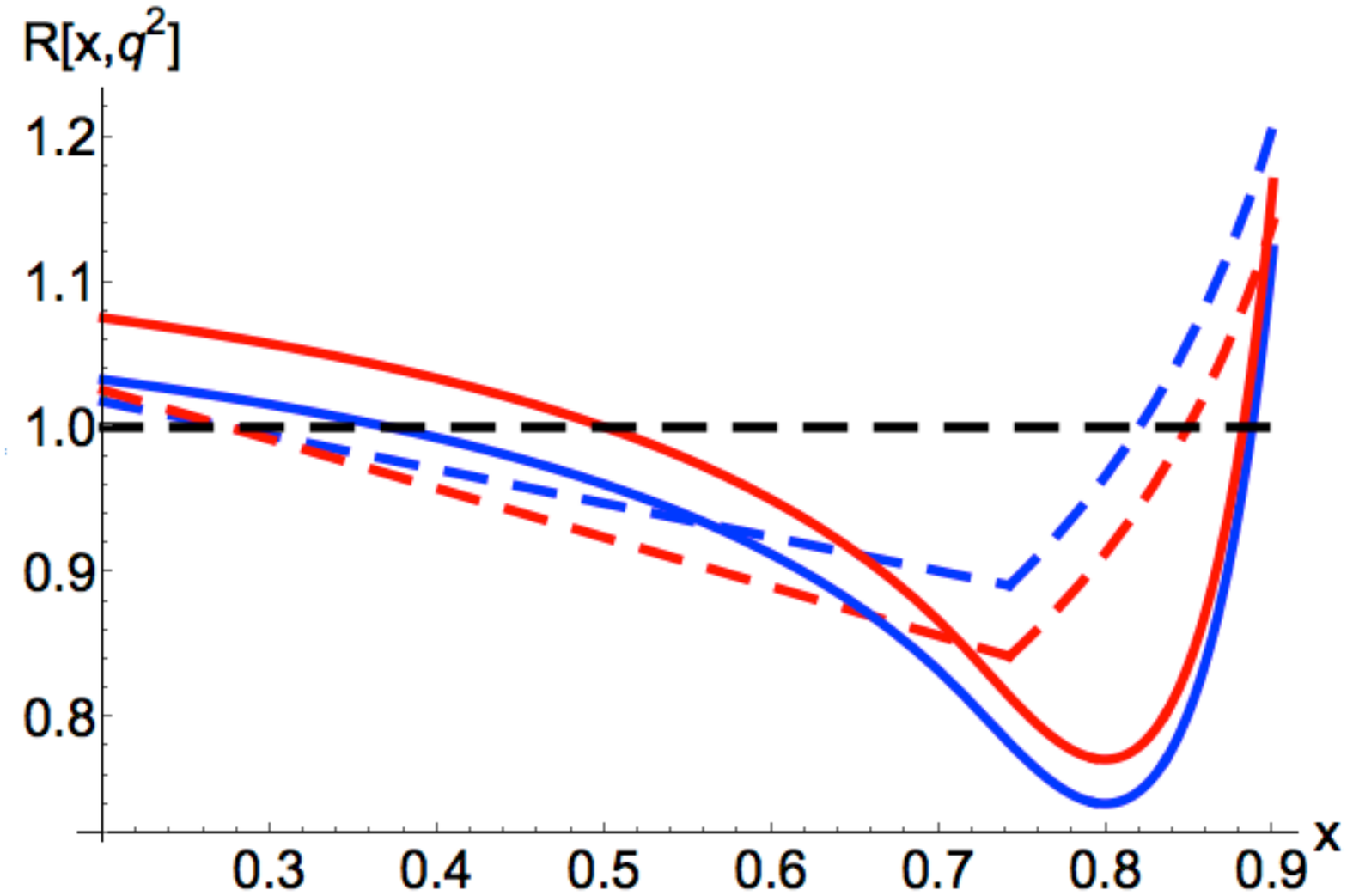}

&
\includegraphics[width=6cm]{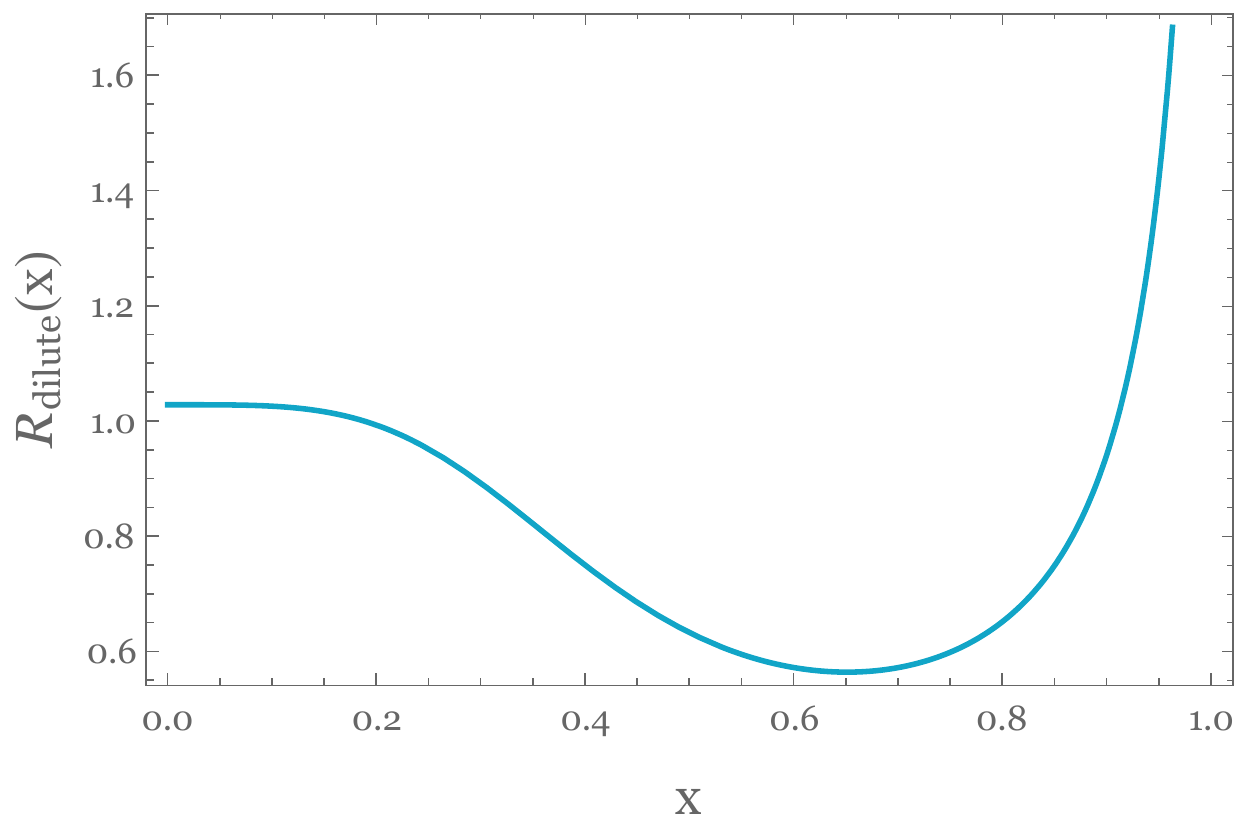}

\\
(a) & (b)
\end{tabular}
\caption{(a) R-ratio at large-x using the leading density contribution and the holographic nucleon structure function
  (solid curves), versus the parametrized empirical ratio (dashed curves), for atomic number $A=12$ (blue curves) and $A=42$ (red curves). (b) Dilute R-ratio for $k_F/m_N=1$, $e_R=0.3$ (U(1)-charge of the bulk fermion), $2\pi^2c_5/\sqrt{4\pi\lambda}=0.01$ (strong coupling), $\tau=3$ (hard scaling exponent),
$j=0.08$ (Pomeron intercept).}\label{dilute-ratio}
\end{center}
\end{figure}

In our bottom-up approach to DIS on a dilute nucleus, we consider DIS on a nucleus
described using a density expansion where the leading density contribution to the current-current correlation of a nucleus is \cite{Mamo:2018eoy}
\begin{eqnarray}
\label{W11}
\frac{{\cal G}_A^{\mu\nu}}{\left<P_A|P_A\right>}
&\approx & \rho_0\frac {4\pi} 3 R_A^3\int\, \frac{d^3p}{2V_3E_p}\,\frac{\theta(p_F-|\vec p|)}{\frac 43 \pi p_F^3}\,
{\cal G}^{\mu\nu}_p\nonumber\\
&+&16\pi\int_{R_A}^{R_A+\Delta}r^2dr\int\, \frac{d^3p}{(2\pi)^3}\frac 1{2V_3E_p}{\theta(p_F(r)-|\vec p|)}\, {\cal G}^{\mu\nu}_p\,,
\end{eqnarray}
where ${\cal G}_p^{\mu\nu}$ is the current-current correlation function for DIS  scattering on a single nucleon.
The R-ratio (the structure function of the dilute nucleus normalized by the structure
function of each constituent nucleon) is plotted in Fig.~\ref{dilute-ratio}a.

\subsection{Top-down Approach to Dilute Nucleus}

In our top-down construction for DIS on a dilute nucleus, we directly do DIS on a
gas of bulk Dirac fermions (protons) undergoing Fermi motion with momentum $k$. In
other words, we compute the one-loop correction to the current-current correlation
function in AdS where the bulk Dirac fermions are running in the loop forming a
Fermi gas. In the large-x regime, we find \cite{Mamo:2019jia}
\begin{equation}
\frac{F_{2}(x,q^2)}{A}\approx 8\pi^2(\tau-1)^2e_R^2\Big(\frac{\beta m_N^2}{q^2}\Big)^{\tau-1}
\,x_{{F}}^{\tau+1}(1-x_{{F}})^{\tau-2}\,,
\end{equation}
where $x_F = \frac{xm_N}{E_F}$ is defined in terms of the Fermi energy $E_F=(k_F^2+m_N^2)^{\frac 12}$ of a single
nucleon with Fermi momentum $k_F$ . Similarly for small-x we have
\begin{equation}
\frac{F_{2}(x,q^2)}{A}\approx {\pi C_{\lambda}}
\Big(\frac{\beta m_N^2}{q^2}\Big)^{\tau-1}\,\frac{1}{x_{{F}}}[I_{0,2\tau+3} + I_{1,2\tau+3}]\,.
\end{equation}
We have plotted the R-ratio in Fig.~\ref{dilute-ratio}b.

\section{DIS on a Dense Nucleus}
We model DIS on a dense nucleus by the corresponding DIS on an extremally
charged (RN)-AdS black hole.

\subsection{DIS on a Classical RN-AdS Black Hole}
\begin{figure}[t]
\begin{center}
\begin{tabular}{cc}
\includegraphics[width=6cm]{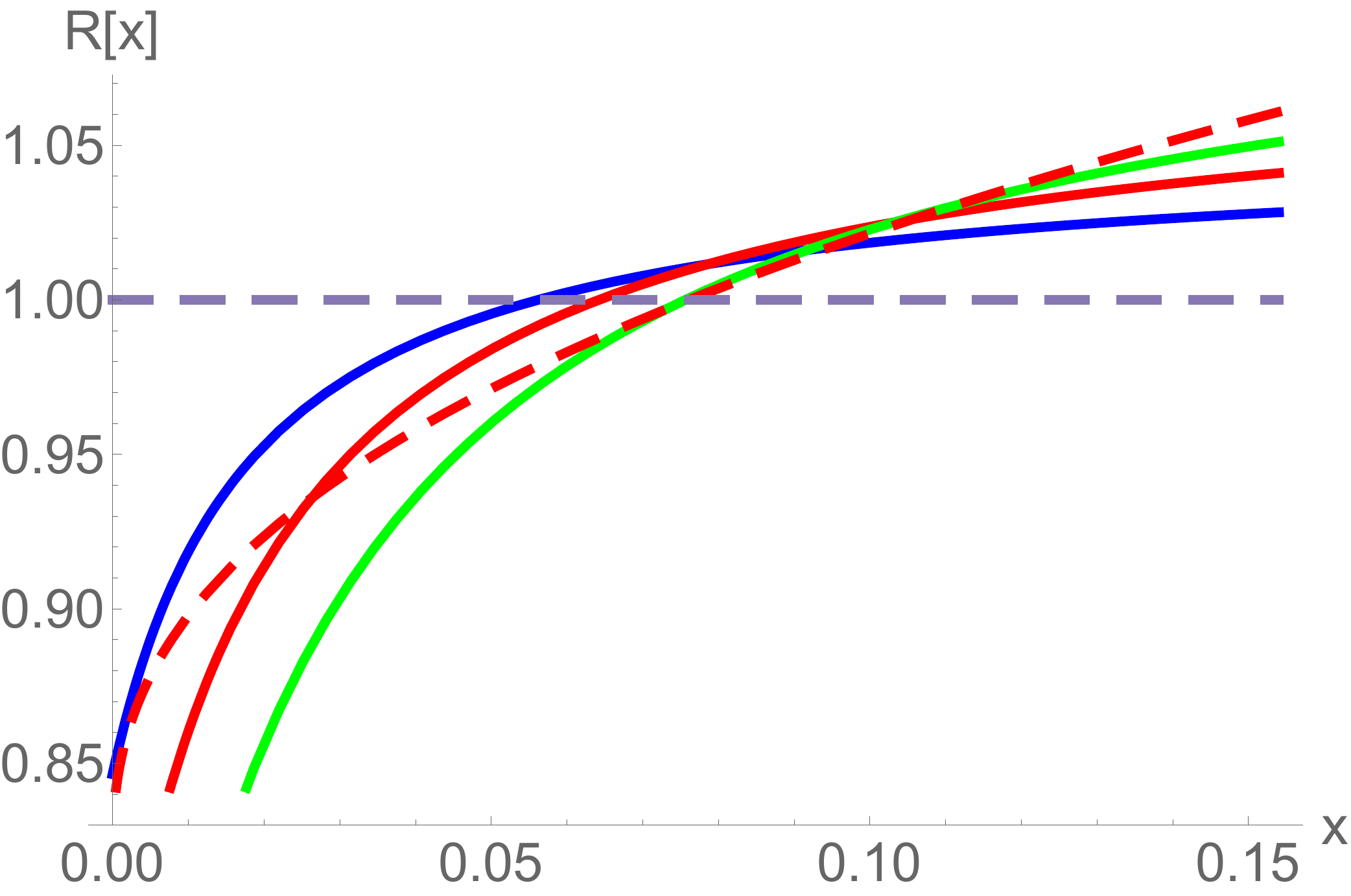}

&
\includegraphics[width=6cm]{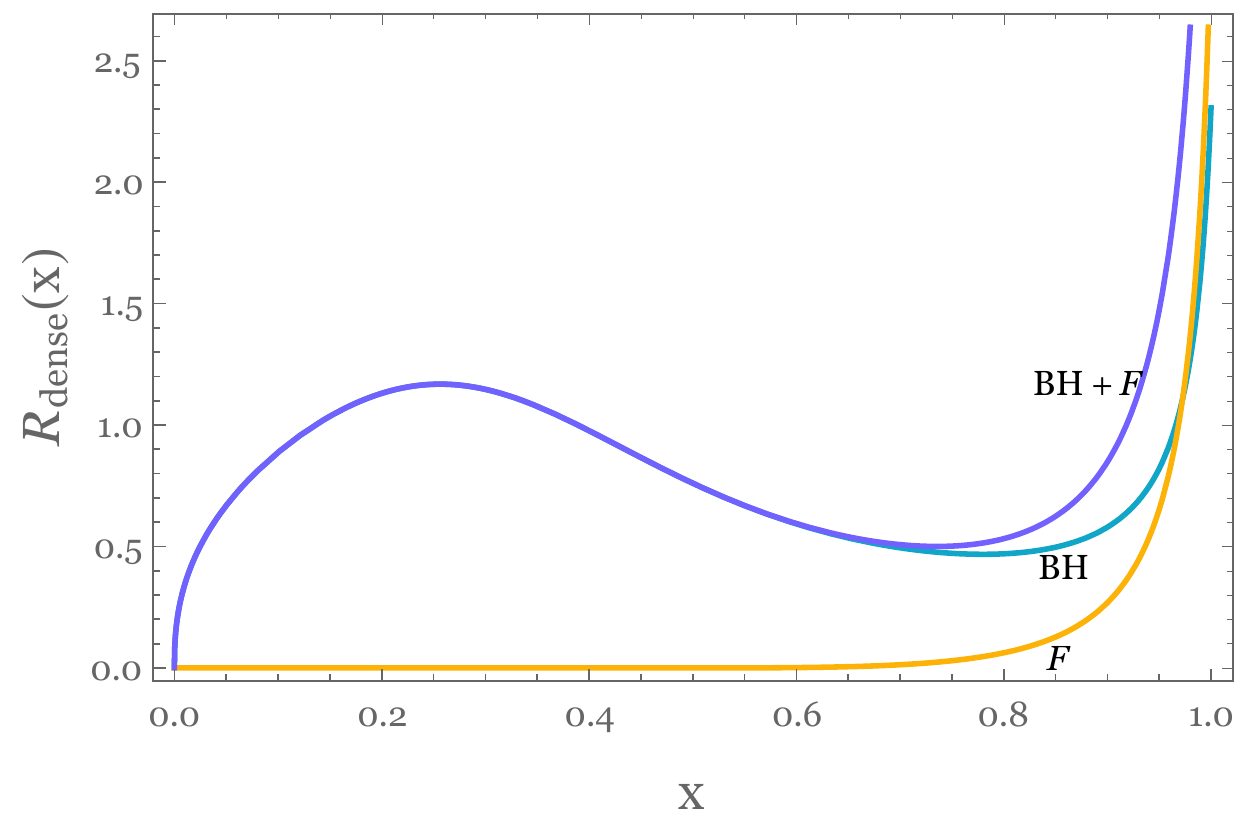}

\\
(a) & (b)
\end{tabular}
\caption{(a) Parametrized DIS data on nuclei (solid curves)  vs holography (dashed curve) in the shadowing region.  (b) Dense R-ratio for $k_F/m_N=3.5$, BH $\equiv$ black hole, F $\equiv$ quantum fermions.}\label{dense-ratio}
\end{center}
\end{figure}

The structure function extracted from DIS on a classical extremal RN-AdS black
hole is given by $F^A_2(x,Q^2) = F_T(x,Q^2) + F_L(x,Q^2)$ where (computed in \cite{Mamo:2018ync}, by closely following \cite{Hatta:2007cs} for uncharged thermal AdS black holes)
\begin{equation}
  \label{NEW}
F^A_T(x,Q^2)=\tilde C_{T}\,\frac{A}{x}\,
\left(\frac{3x^2Q^2}{4m^2_N}\right)^{\frac 23}\,,\,\,\,\,\,
F^A_L(x,Q^2)=\tilde C_{L}\,\frac{3A}{4x}
\left(\frac{3x^2Q^2}{4m^2_N} \right)\,,
\end{equation}
with $\tilde C_{T,L}/C_{T,L}=\pi^5{(48\alpha)^2}/{2N_{c}^2}$. We have plotted the R-ratio in Fig.~\ref{dense-ratio}a.

\subsection{DIS on a Quantum RN-AdS Black Hole}
The leading order structure function (in $1/N_c$ expansion of the current-current correlation
function) due to the classical RN-AdS black hole receives one-loop correction from the quantum fermions hovering around the extremal RN-AdS black hole. Therefore, by including the quantum correction due to the bulk Dirac fermions
(protons), the total structure function of the dense nucleus, modeled as a quantum RN-AdS black hole is given by\cite{Mamo:2019jia}
\begin{eqnarray}
\frac{F_2^{\rm dense}(x,q^2)}{A}&\approx & C_T\left(\frac{3q^2}{4m_N^2}\right)^{\frac 23}\,x^{\frac 13}+C_{AdS2}\,e_R^2\,\left(\frac{\mu^2}{q^2}\right)^{\nu_{k_F}+2}\nonumber\\
&\times & x_{k_F}^{\nu_{k_F}+5}(1-x_{k_F})^{\tau-\frac 32}\,{}_2F_1^2\left(\tau_+, \tau_-, \tau-1, 1-x_{k_F}\right)\,,
\end{eqnarray}
where the first contribution (which is the leading and dominant contribution) stems
from DIS on the classical black hole, and the second and subleading contribution
stems from DIS scattering of the emerging holographic Fermi liquid near the black
hole horizon which is a quantum correction that is vanishingly small at small-x.
We have plotted the R-ratio in Fig.~\ref{dense-ratio}b.

\section{Summary and Conclusion}
We have shown that the quantum extremal RN-AdS black hole exhibits shadowing in small-x regime, see Fig.~\ref{dense-ratio}, and EMC effect in the large-x regime where it is dominated by a Fermi gas in AdS spacetime, see Fig.~\ref{dilute-ratio}.

\section{Acknowledgments}
I thank Ismail Zahed for collaboration on this work. This work was supported by the U.S. Department of Energy under Contract No. DE-FG-88ER40388.


\begin{thebibliography}{10}

\bibitem{lamp94}
L.~Lamport, {\em \LaTeX, A Document Preparation System}, 2nd edn.
  (Addison-Wesley, Reading, MA, 1994).

\bibitem{ams04}
\AmS, {\em \AmS-\LaTeX{} Version 2 User's Guide} (American Mathematical
  Society, Providence, 2004), \url{http://www.ams.org/tex/amslatex.html}.

\bibitem{jarl88}
C.~Jarlskog, {\em CP {V}iolation} (World Scientific, Singapore, 1988).

\bibitem{best03}
B.~W. Bestbury, {$R$}-matrices and the magic square, {\em J. Phys. A} {\bf 36},
  1947  (2003).

\bibitem{pier02}
P.~X. Deligne and B.~H. Gross, On the exceptional series, and its descendants,
  {\em C. R. Math. Acad. Sci. Paris} {\bf 335}, 877  (2002).

\bibitem{jame02}
J.~M. Landsberg and L.~Manivel, Triality, exceptional {L}ie algebras and
  {D}eligne dimension formulas, {\em Adv. Math.} {\bf 171}, 59  (2002),
  \url{http://www.url.com/triality.html}.

\bibitem{weis94}
G.~H. Weiss (ed.), {\em Contemporary {P}roblems in {S}tatistical {P}hysics}
  (SIAM, Philadelphia, 1994).

\bibitem{gupt97}
R.~K. Gupta and S.~D. Senturia, Pull-in time dynamics as a measure of absolute
  pressure, in {\em Proc. {IEEE} Int. Workshop on Microelectromechanical
  Systems ({MEMS}'97)\/},  (Nagoya, Japan, 1997).

\bibitem{rich60}
L.~F. Richardson, {\em Arms and {I}nsecurity} (Boxwood, Pittsburg, 1960).

\bibitem{chur90}
R.~V. Churchill and J.~W. Brown, {\em Complex {V}ariables and {A}pplications},
  5th edn. (McGraw-Hill, 1990).

\bibitem{benh93}
F.~Benhamou and A.~Colmerauer (eds.), {\em Constraint {L}ogic {P}rogramming,
  {S}elected {R}esearch} (MIT Press, 1993).

\bibitem{bake72}
D.~W. Baker and N.~L. Carter, {\em Seismic {V}elocity {A}nisotropy {C}alculated
  for {U}ltramafic {M}inerals and {A}ggregates}, in {\em Flow and {F}racture of
  {R}ocks\/},  eds. H.~C. Heard, I.~V. Borg, N.~L. Carter and C.~B. Raleigh,
  Geophys. Mono., Vol.~16 (Am. Geophys. Union, 1972), pp. 157--166.

\bibitem{hobb92}
J.~D. Hobby, {\em {A User's Manual for MetaPost}}, Tech. Rep. 162, AT\&T Bell
  Laboratories (Murray Hill, New Jersey, 1992).

\bibitem{bria84}
B.~W. Kernighan, {\em {PIC}---{A} Graphics Language for Typesetting}, Computing
  Science Technical Report 116, AT\&T Bell Laboratories (Murray Hill, New
  Jersey, 1984).

\bibitem{hear94}
H.~C. Heard, I.~V. Borg, N.~L. Carter and C.~B. Raleigh, {VoQS: Voice Quality
  Symbols}, Revised to 1994,  (1994).

\bibitem{brow88}
M.~E. Brown, An interactive environment for literate programming, PhD thesis,
  Texas A\&M University, (TX, USA, 1988), pp. ix + 102.

\bibitem{lodh74}
G.~S. Lodha, Quantitative interpretation of ariborne electromagnetic response
  for a spherical model, Master's thesis, University of Toronto  (1974).

\bibitem{dani73}
D.~Jones, {The term `phoneme'}, in {\em Phonetics in Linguistics: A Book of
  Reading\/},  eds. W.~E. Jones and J.~Laver (Longman, London, 1973) pp.
  187--204.

\bibitem{davi93}
B.~Davidsen, Netpbm  (1993),
  \url{ftp://ftp.wustl.edu/graphics/graphics/packages/NetPBM}.

\end{thebibliography}


\begin{thebibliography}{99}
\bibitem{Mamo:2018eoy}
  K.~A.~Mamo and I.~Zahed,
  Phys.\ Rev.\ D {\bf 100}, no. 4, 046015 (2019)
\bibitem{Mamo:2018ync}
  K.~A.~Mamo and I.~Zahed,
  arXiv:1807.07969 [hep-th].

\bibitem{Mamo:2019jia}
  K.~A.~Mamo and I.~Zahed,
  arXiv:1905.07864 [hep-th].

\bibitem{Basar:2017ocn}
  G.~Basar, D.~E.~Kharzeev, H.~U.~Yee and I.~Zahed,
  Phys.\ Rev.\ D {\bf 95}, no. 12, 126005 (2017); H.~U.~Yee, this proceeding.

\bibitem{Polchinski:2002jw}
  J.~Polchinski and M.~J.~Strassler,
  JHEP {\bf 0305}, 012 (2003)


\bibitem{Hatta:2007cs}
  Y.~Hatta, E.~Iancu and A.~H.~Mueller,
  JHEP {\bf 0801}, 063 (2008)

\end{thebibliography}
\end{document}